# Three-dimensional photonic topological insulator induced by lattice dislocations

Eran Lustig[†,1], Lukas J. Maczewsky[†,2], Julius Beck[2], Tobias Biesenthal[2], Matthias Heinrich[2], Zhaoju Yang[3], Yonatan Plotnik[1], Alexander Szameit[2] and Mordechai Segev[1,*]

[1] *Physics Department and Solid State Institute, Technion-Israel Institute of Technology, Haifa 3200003, Israel*

[2] *Institut für Physik, Universität Rostock, Albert-Einstein-Str. 23, 18059 Rostock, Germany*

[3] *Department of Physics, Zhejiang University, Hangzhou 310058, China*

[†] *These authors contributed equally to this work*

*msegev@technion.ac.il

**The hallmark of topological insulators is the scatter-free propagation of waves in topologically protected edge channels**[1–3]**. This transport is strictly chiral on the outer edge of the medium, and therefore capable of bypassing sharp corners and imperfections, even in the presence of substantial disorder. In photonics, two-dimensional topological edge states have been demonstrated on several different platforms**[4–6]**, and are emerging as a promising tool for robust lasers**[7–10]**, quantum devices**[11–14]**, and other applications. However, three-dimensional photonic topological insulators, specifically those supporting topologically protected edge states in all 3D, have thus far remained out of experimental reach. Here, we demonstrate a three-dimensional photonic topological insulator with protected topological edge states. The topological protection is enabled by a screw dislocation. For this purpose, we utilize the concept of synthetic dimensions**[15–18] **in a 2D photonic waveguide array**[19] **by introducing an additional modal dimension to transform the system into a 3D photonic topological insulator**[20]**. The lattice dislocation endows the system with edge states propagating along three-dimensional trajectories, with topological protection akin to strong photonic topological insulators**[3,21,22]**. Our work paves the way for utilizing three-dimensional topology in photonic science and technology.**

Photonic topological insulators (TIs) are systems that facilitate a robust and unidirectional flow of light along the edges of the device[4–6,23]. Based on similar principles as electronic TIs[1–3], these artificial electromagnetic media are engineered to exhibit a topologically non-trivial band structure, and constitute a promising platform for applications ranging from quantum computing[24] to spintronics[25] as well as a means of forcing extended ensembles of laser emitters to act as one laser [7–10], and various applications in quantum optics[11–14]. However, unlike electronic TIs, topological photonics has thus far mostly relied on 1D[26,27] and 2D geometries - essentially confining them to a small subset of possible topological phases. Since photons propagate at a great speed and only interact weakly with surrounding fields, realizing 3D TIs for photons has remained a formidable challenge.

Three-dimensional TIs that obey time-reversal (TR) symmetry are generally divided into two categories: weak and strong [28–31]. Strong TIs host 3D edge states on all of their surfaces, and are impervious to variations in the shape of the medium or variation that are small compared to the band gap energy. In contrast, weak 3D TIs are topologically equivalent to stacked arrangements of 2D TIs, together forming a 3D topological crystal. In a similar vein, systems lacking TR symmetry may also support 3D TIs by stacking 2D TIs. These too are considered "weak" as they do not exhibit topologically protected surface states[22]. For example, a 3D cubic lattice with a constant magnetic field along one of the lattice axes is known to form a 3D weak TI (Fig.1a). Since each of the 2D layers supports an edge state, the 3D composite structure is characterized by multiple edge states propagating on four surfaces in a 2D course (Fig.1b).

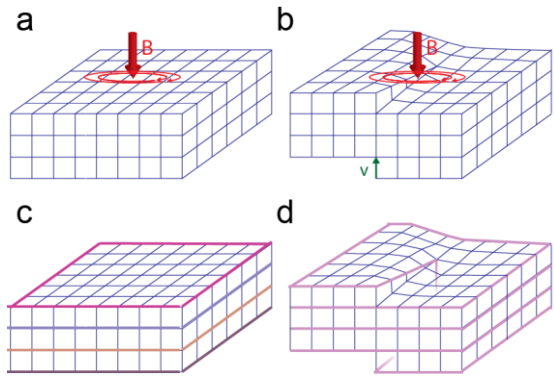

**Figure 1: A weak TI vs a weak TI with a dislocation. (a)** Illustration of a cubic lattice with a constant magnetic field along one of its axes, that maps to stacking layers of 2D TIs. **(b)** Edge states from the different square TIs that can couple to each other and form a gap. **(c)** A weak TI (from (a)) with a dislocation defined by the Burgers vector $v$. **(d)** In the presence of a screw dislocation, all individual edge state in a weak TI merge into edge states that wind around the entire 3D system. These edge states do not map to degenerate edge states of a weak TI anymore, but instead give rise to a topologically protected 3D TI.

Notably, in a weak TI, the individual edge states couple to one another and form a dispersion curve on the surface of the lattice. Consequently, the edge states can in principle form a gap, rendering them vulnerable to disorder[32]. Such structures, were recently suggested [33–36] and demonstrated [37] in magneto-optic materials at microwave frequencies. An interesting situation occurs when a screw dislocation is introduced into such a system (Fig.1c) [21]. In contrast to the coupling between the planar topological edge states in the pure lattice of a weak TI, the entirety of all the edge states merges into a single edge channel that winds helically around the outer surface of the 3D system. As a result, the phases between neighboring edges are strictly fixed, preventing the formation of a band gap and rendering the transport immune to disorder (Fig.1d). In other words, the dislocation forces the edge states to propagate in all three dimensions and endows the 3D weak TI with topological protection to its edge states[22,38]. Recently, such dislocations were demonstrated in acoustic[39,40], mechanical[41], and electronic systems[42,43].

In photonics, where magnetic interactions are prohibitively weak at optical frequencies, unorthodox approaches are required to tackle the problem of implementing the physics of TIs on electromagnetic waves [5,6,44]. Recently, the concept of synthetic dimensions has gained popularity for exploring effects that are otherwise unapproachable due to limitations in geometry, connectivity and fields in real space [15,45–47]. In photonics, effects requiring three or more dimensions, such as Thouless pumping in high dimensions, Weyl points, disorder in high dimensions and other effects, were successfully demonstrated by reinterpreting certain parameters of the system's Hamiltonian as spatial coordinates[46,48–51] or by using fiber loops to instantiate time-discrete quantum walks[52]. Alternatively, the use of coupled modal ladders was shown to allow exploring the full dynamics in synthetic space [15,53,54]. Using this modal ladder technique, photonic TIs[16–18] were demonstrated in hybrid lattices with one spatial axis and one modal axis[19,55]. To date however, experimental realizations of 3D TIs at optical frequencies remain elusive, and more generally – topologically protected edge states propagating in three dimensions have never been observed with electromagnetic waves.

Here, we demonstrate 3D photonic TIs that support topologically protected edge states propagating in all three dimensions by virtue of a dislocation, and present the first realization of a Floquet 3D TI. Hence, this work paves the way for both the study of high-dimensional structures in photonics, and the interplay of dislocations with topology in general lattice systems. To implement the 3D TI with a dislocation, we employ waveguide lattices with two spatial, $(x, y)$, and one modal dimensions. In this configuration, the third spatial coordinate, $z$, plays the role of time [5,56,57] The evolution of the light is governed by the paraxial wave equation:

$$i\partial_z \psi(x,y,z) = -\frac{1}{2k_0}\nabla^2 \psi(x,y,z) - \frac{k_0 \Delta n(x,y,z)}{n_0}\psi(x,y,z)$$

where $\psi$ is the field, $k_0$ is the wavenumber in vacuum, $n_0$ is the ambient refractive index, and $\Delta n$ is the local variation in refractive index that forms the waveguides in our system. To explain how our waveguide structure implements a 3D TI with a dislocation, we will first describe a single 2D layer lacking dislocations. An individual 2D layer is comprised of waveguides in a square lattice with lattice constant $a$ (Fig.2a). Each unit cell in this square lattice includes two waveguides, which rotate around their mean transverse coordinates with a spatial frequency $\Omega$ along the propagation axis $z$, but with a relative phase of $\pi/2$. The lattice is therefore comprised of two sublattices of helical waveguides with a relative phase difference of $\pi/2$. Such a structure was shown to be a 2D anomalous Floquet TI [58–60].

In order to introduce the desired additional dimension, we replace each waveguide in Fig.2a with a column of waveguides, differing only in their respective effective index of refraction (Fig.2b-c). This manifests a Stark ladder of modes (Fig.2d) that, in the following, will serve as the synthetic dimension. The overall structure in real space is shown in Fig.2e. Crucially, while any column interacts with its neighboring ones, each mode of each column couples only to a mode with similar wavenumber $k_z$ due to phase-matching. Thus, in the entire lattice, each mode functions as a 2D layer that is, for all intents and purposes, disconnected from other modes (other 2D layers). Furthermore, the parameters $(\Omega, a)$ are chosen such that the 2D lattice of each mode is in the topological phase and is capable of supporting a topological edge state. Consequently, our structure constitutes a 3D weak TI in synthetic dimensions (Fig.2f) that, before we introduce the dislocation, maps to stacked layers of 2D TIs. To introduce the dislocation, we introduce inter-mode coupling along a line stretching from the lower edge to the middle of the lattice. For the inter-mode coupling to be as efficient as the regular intra-mode coupling, we shift the oscillation path and modulate the refractive index, such that at the region of interaction the appropriate waveguides will be phase matched and in proximity (see Methods).

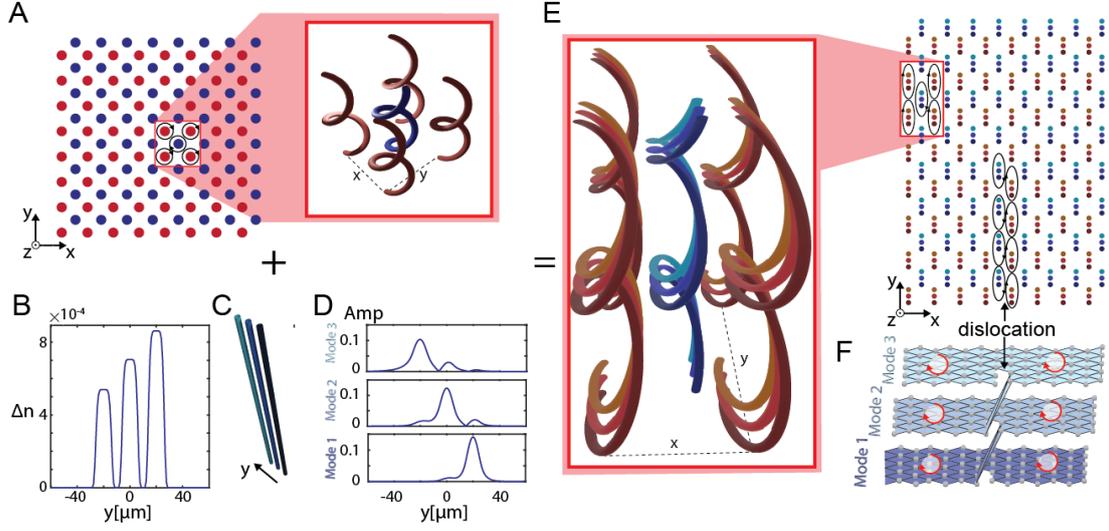

**Figure 2**: **Weak TI with photonic waveguide arrays**. **(a)** The anomalous quantum Hall effect in a waveguide array as was experimentally demonstrated in [59,60]. Each unit cell has 2 helical waveguides rotating with a phase difference of π/2 between them. **(b,c)** Gradually-increasing refractive index contrast in each column of 3 waveguides as a function of position along the column, as indicated by the graded blue colors **(d)** Shape of the localized modes induced by the index gradient in (b). **(e)** Lattice constructed by replacing each waveguide from (a) with a 3-waveguide column from (b-c). The black ellipses represent the location of the center of the columns in the modulation trajectory where in the dislocation the trajectory is shifted **(f)** Synthetic-space diagram of the lattice in (e), highlighting its correspondence to a 3D weak TI. The synthetic-space lattice contains layers of 2D TIs for each mode. The marked dislocation converts the 3D weak TI, to a 3D TI with a dislocation.

Unlike ordinary weak 3D TI, the edge states in our structure are topologically protected in all 3 dimensions[38]. The triad of topological invariants associated with our structure is (0,0,1), where the first two invariants are associated with the spatial dimension, and the last with the mode dimension, signifying a non-trivial Rudner number[58]. Since the Burgers vector (of the dislocation) points in the direction of the mode ladder and its length is one hopping in the mode ladder, the dislocation induces topological protection on the surface states of the 3D TI[61,62] (see supplementary information). Accordingly, as shown in Fig. 3a, when our waveguide array is devoid of dislocations, the resulting edge states are degenerate and remain strictly confined to individual layers in synthetic space. This can be viewed in Fig.3c, where the emerging edge states form degenerate groups of states. In this case, there is no topological protection from localization and other disorder-induced dispersion effects in the mode direction, and a gap can be formed by inducing various perturbations [32].

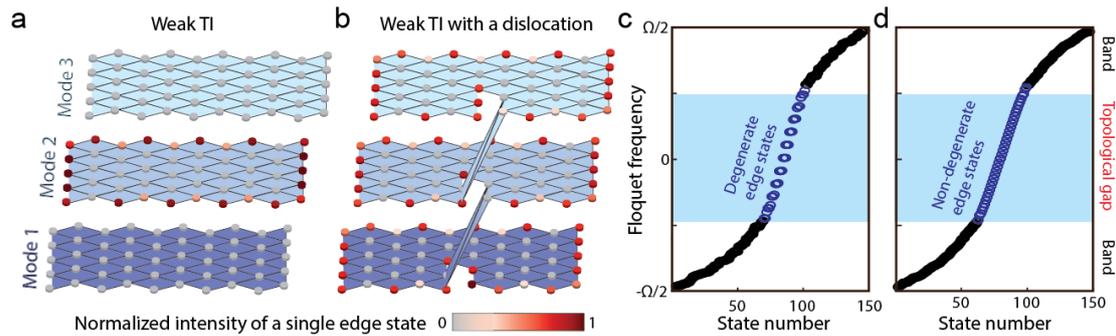

**Figure 3**: **3D TI with a dislocation. (a)** Formation of an edge state of the 3-mode synthetic-space lattice from Fig.2e without the dislocation (using the experimental parameters in a 5 by 10 lattice). Each mode has a set of 2D edge states of its own. Since the layers are uncoupled, each edge state belongs to a different layer, highlighting that the structure is a weak 3D TI. **(b)** Edge states of the weak 3D TI with a dislocation (the lattice of Fig.2e), formed by the localized coupling between adjacent mode at a specific position. The edge-states are extended over the entire three-dimensional circumference. **(c-d)** Floquet spatial frequency ($k_z$) for the Weak TI in (a) and the Weak TI with a dislocation in (b), calculated with our experimental parameters. Notice that in (c) the edge states are degenerate (each edge state is in a different layer) whereas in (d) there is no degeneracy.

Introducing a dislocation into this structure changes the situation profoundly. Figures 3b,d show how the presence of the dislocation impacts the structure of the edge states and their respective eigenenergies: The degeneracy is lifted, and the resulting dispersive branch bridges diagonally the gap between the bulk bands. Figure 3b also illustrates the path that an edge wave packet will take: Starting at the top layer and moving in clockwise orientation, an initial excitation will descend to the layer below it each time it encounters the dislocation. Finally, after the lowest modal layer is reached, the wavepacket ascends along the dislocation axis back to the highest mode, thereby completing a genuine three-dimensional loop. In this case the edge states are topologically protected, and thus will not be impaired by inter-layer coupling, which naturally exists in our system due to the modulation

We experimentally realize this phenomenon by fabricating the waveguide structure sketched in Fig.2e using the femtosecond laser direct writing technique[63]. To trace the propagation of the light along its protected 3D trajectory, we launch a 633nm wavelength laser beam by using a spatial light modulator [64] on the edge of the structure, and observe the intensity distribution after 15cm of propagation at the output facet. We are aiming to show how the light is propagating in a topologically protected fashion in all three dimensions – the two spatial dimensions and the synthetic modal dimension. The propagation of the edge state in the 2D layers is demonstrated with three excitations (one in each of the modes) in the unperturbed region of the lattice (Fig. 4a-c). Since the modes are tightly localized, the position in modal space can be chosen by injecting light into one of the three waveguides. Far away from the dislocation, we indeed observe that the initial excitations remain in their initial modes as they propagate along the edge and around a corner, as shown in the intensity pictures at the output facet of the waveguide array (Fig.4d-f). For example, in Fig.4c we excite a wavepacket on the edge that is entirely in the third mode, by exciting only the lowest waveguide in four columns along the upper edge. Since the modes are localized, we can directly identify the mode from observing the output intensity in the corresponding Fig.4f. Accordingly, the light at the output does not couple to other modes (i.e. it remains at the lower waveguide of each column) and continues to propagate along the edge bypassing the corner.

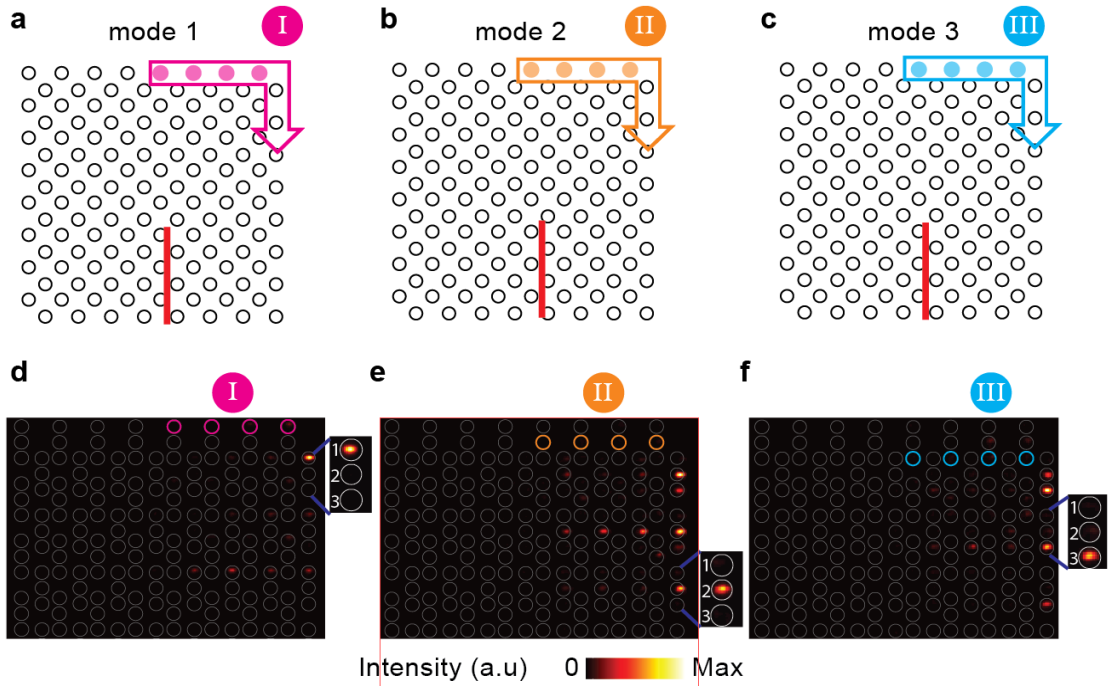

**Figure 4: Experimentally viewed evolution of edge wavepackets in the 3D synthetic-space TI. (a-c)** Sketches showing the excitation in modal space of modes 1-3 respectively. The full circles are the excited waveguides in each of the modes, and the arrow points to the propagation direction in the topological phase. **(d-f)** Intensity images at the output facet of the structure. The painted circles mark the areas into which light is injected at the input. The light propagates away from the excitation region, passes the corner, and continues along the next (vertical) edge. At the output facet, the light is localized in each column in one of the 3 waveguides, where the uppermost waveguide of the column represents mode 1, and the lowest waveguide is mode 3. The mode is determined according to the position of the light in the column. A colored number ($I, II, III$) is attributed to each excitation.

It is instructive to view the propagation in modal space near the intersection of the dislocation and the edge (Fig. 5), where we excite the light near the dislocation, to observe the evolution in the modal dimension. Unlike the excitation presented in Fig.4, here, light injected into a certain mode and descends to the mode below it upon passing the dislocation ((b) and (c) in Fig.5a-c). The exception is when the beam is launched in the first mode and has no layer to descend to ((a) in Fig.5a). In this case, the dislocation behaves as a barrier and the light will flow around it (along the dislocation) without scattering, staying in the lowest mode. As in Fig.4a-c, Fig.5a-c show the excitations and the evolution in 3D space, but unlike in Fig.4a-c, here the dislocation serves to transfer the light between different modes. We track each excitation with the colored numbers ($I, II, III$).

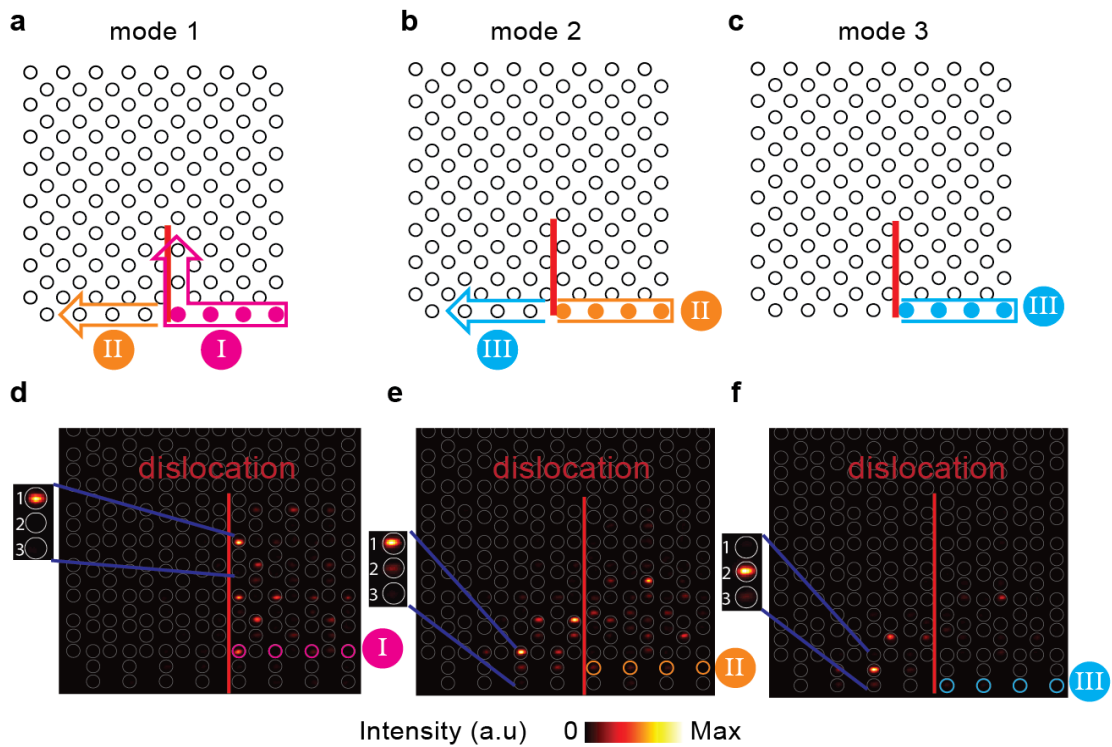

**Figure 5**: **Experimentally viewed edge wavepackets bring transformed from each mode to the mode below it by the intersection of the edge and the dislocation**. The markings are similar to Fig.4, but here the excitations are on the edge near the dislocation. **(a-c)** Sketches of the mode transfer driven by the dislocation. The full circles mark the excited sites and the arrow points to the propagation direction in the topological phase. **(d-f)** Intensity images at the output facet of our waveguide array. The colored centers mark the waveguides into which light is injected at the input. The mode is determined by the position of the light in the column. A colored number ($I, II, III$) is attributed to each excitation.

Figure 5d-f shows the experimentally viewed output intensity. As in Fig.4 the mode can be identified by the position of the localized light in each column. For example, when launching a wavepacket in mode 2 (light localized on the middle waveguide of each column), near the dislocation, marked by the orange number (II) in Fig.5 a,b, the light propagates toward the dislocation and then descends to mode 1 (light localized on the upper waveguide of each column) at the other side of it (Fig.5e), continuing to propagate on the 3D edge. On the other hand, when the light is launched at mode 1 in the same location (Fig.5a), the light simply bypasses the dislocation and stays in the same mode – at the upper waveguide (Fig.5.d). Our observations (Fig. 5d-F) unequivocally show that the topological edge state indeed follows a trajectory in all three dimensions.

In conclusion, we experimentally investigated the dynamics of edge states in a weak 3D photonic TI, and showed that the introduction of a screw dislocation endows the system with properties of a strong 3D TI. This is the first observation of a 3D photonic TI, and also the first 3D TI in synthetic space in any system. We expect that this work will open the door for exploring higher-dimensional topological phases in laboratory experiments.

Acknowledgements
The authors would like to thank C. Otto for preparing the high-quality fused silica samples used for the inscription of all photonic structures employed in this work. The Technion team gratefully acknowledges the support of an Advanced Grant from the European Research Council (ERC) under the European Union's Horizon 2020 research and innovation program (grant agreement no. 789339), and the support of a research grant from the Air Force Office of Scientific Research (AFOSR) of the USA. The Rostock team gratefully acknowledges the support the Deutsche Forschungsgemeinschaft (grants SCHE 612/6-1, SZ 276/12-1, BL 574/13-1, SZ 276/15-1, SZ 276/20-1 and SFB 1477 "Light-Matter Interactions at Interfaces", project number 441234705) and the Alfried Krupp von Bohlen and Halbach Foundation.